# Vector exceptional points with strong superchiral fields

Tong Wu[1*], Weixuan Zhang[1*], Huizhen Zhang[1*], Saisai Hou[1], Guangyuan Chen[1], Ruibin Liu[1], Cuicui Lu[1], Jiafang Li[1], Rongyao Wang[1], Pengfei Duan[2], Junjie Li[3], Bo Wang[4], Lei Shi[4], Jian Zi[4] and Xiangdong Zhang[1$]

[1]*Key Laboratory of advanced optoelectronic quantum architecture and measurements of Ministry of Education, Beijing Key Laboratory of Nanophotonics & Ultrafine Optoelectronic Systems, School of Physics, Beijing Institute of Technology, 100081, Beijing, China;*

[2]*CAS Center for Excellence in Nanoscience, CAS Key Laboratory of Nanosystem and Hierarchical Fabrication, National Center for Nanoscience and Technology (NCNST), Beijing 100190, China.*

[3]*Beijing National Laboratory for Condensed Matter Physics, Institute of Physics, Chinese Academy of Sciences, Beijing 100190, China*

[4]*State Key Laboratory of Surface Physics, Key Laboratory of Micro- and Nano-Photonic Structures (Ministry of Education) and Department of Physics, Fudan University, Shanghai 200433, China*

*These authors contributed equally to this work.*

[$]*Author to whom any correspondence should be addressed: zhangxd@bit.edu.cn*

## Abstract

Exceptional points (EPs)，branch points of complex energy surfaces at which eigenvalues and eigenvectors coalesce, are ubiquitous in non-Hermitian systems. Many novel properties and applications have been proposed around the EPs. One of the important applications is to enhance the detection sensitivity. However, due to the lack of single-handed superchiral fields, all of the proposed EP-based sensing mechanisms are only useful for the non-chiral discrimination. Here, we propose theoretically and demonstrate experimentally a new type of EP, which is a called radiation vector EP, to fulfill the homogeneous superchiral fields for chiral sensing. This type of EP is realized by suitably tuning the coupling strength and radiation losses for a pair of orthogonal polarization modes in the photonic crystal slab. Based on the unique modal-coupling property at the vector EP, we demonstrate that the uniform superchiral fields can be generated with two beams of lights illuminating on the photonic crystal slab from opposite directions. Thus, the designed photonic crystal slab, which supports the vector EP, can be used to perform surface-enhanced chiral detection. Our findings provide a new strategy for ultrasensitive characterization and quantification of molecular chirality, a key aspect for various bioscience and biomedicine applications.

Chirality plays a crucial role in modern biochemistry and the evolution of life [1, 2]. Many biologically active molecules are chiral, detecting and characterizing chiral enantiomers of these biomolecules are of considerable importance for biomedical diagnostics and pathogen analyses [3-5]. The molecular chirality discrimination can be realized with circular dichroism (CD) spectroscopy describing the difference in molecular absorption of left- and right-handed circularly polarized lights (CPLs) [6, 7]. In general, the molecular CD signal is typically weak; thus, chiral analyses by such a spectroscopic technique have usually been restricted to analysis at a relatively high concentration [8-11]. Early works proposed by Tang and Cohen [12, 13] have shown that the enantioselectivity of optical excitation of a molecule is highly dependent on the chirality of optical field, and so it stands that the superchiral field, which possesses larger optical chirality than CPLs, can lead to significant enhancement of enantioselective excitation of chiral molecules allowing for ultrasensitive detection and characterization of chiral molecules. Remarkably, the potential of superchiral fields is far more beyond enhancing the difference of absorption between enantimitors, it triggers the development of almost all the spectroscopy technologies related to the optical activity phenomenon, from Raman optical activity, fluorescence CD, to optical rotation. The superchiral field can also enhance the difference between total optical forces on enantiomers, which promises the development of application for chiral molecule sorting. Recent investigation has shown that superchiral field has a potential application on the enantioselective conversion in reactions [14], extending its use from the all-optical areas to photochemistry fields. Because of its high importance in all these applications, the superchiral field has bloomed in the last 10 years, and intense theoretical and experimental studies have been devoted to the generalization of strong superchiral near-fields with the help of artificial nanostructures [15-25]. Despite of these progresses, many schemes suffer from high optical losses [15-25], extrinsic CD caused by nanostructure itself [24] and different handedness of chiral near-fields [21-25]. Thus, how to engineer the achiral nanostructure to produce strong single-handed superchiral fields remains challenging.

On the other hand, in contrast to Hermitian systems, the special degeneracies called exceptional points (EPs) widely exist in the non-Hermitian systems [26-28]. Both the real and imaginary parts of the eigenvalues coalesce at this point. The EPs have been observed in many photonic systems, such as optical microcavities [29, 30], coupled optical waveguides [31-33], photonic crystal slabs [34-37] and unidirectionally coupled resonators [38, 39]. Many novel properties are also found at or near the EPs, such as loss-induced suppression and revival of lasing [40-42], unidirectional transmission or reflection [43-48], topological chirality [49-51] and laser mode selectivity [52-54]. Recently, the photonic nanostructures, which sustain higher-order EPs, have been demonstrated to be achievable and can be used for ultra-sensitive nanoparticle sensing [55-61]. However, these nano-systems, which cannot create single-handed superchiral near-fields, are not suitable for chiral discrimination. It is therefore relevant to ask whether strong superchiral fields can be created and the ultra-sensitive chiral detection can be realized with the help of EPs.

In this Letter, we experimentally demonstrate a new type of EP called radiation vector EP. This type of EP is created by suitably tuning the coupling strength between two orthogonal polarization modes and their radiation losses in the photonics crystal slab. The eigenstate of this vector EP possesses strong and homogeneous superchiral fields. Based on the temporal coupled-mode theory, we find that two orthogonal polarization modes are nearly coupled to the opposite output ports at the vector EP. In this case, the strong superchiral fields can be generated near the photonic crystal slab excited with two beams of lights coming from opposite directions. Thus, our study opens up a new way for ultra-sensitive detecting of molecular chirality.

We consider a photonic crystal slab with a square array of cylindrical holes (periodicity $a$, diameter $d$ and thickness $h$) coated on a substrate, as shown in Fig. 1a. The refractive indexes of the background, photonic crystal slab and substrate are labeled by $n_b$, $n_p$ and $n_s$, respectively. At the Γ point of the Brillouin zone, the system processes $C_{4v}$ symmetry where both one- and two-dimensional irreducible representations exist. Based on the symmetry arguments, only doubly-degenerate Bloch modes can couple with free-space which give rise to radiation losses. We first consider the structure which is symmetric with respect to the $z=0$ plane ($n_b = n_s$). For such a structure, the Bloch modes can be classified into two types with symmetric (TE-like) and anti-symmetric (TM-like) distributions of electric fields. By tuning the diameter of cylindrical hole ($d$) and thickness of the photonic crystal slab ($h$), a pair of doubly degenerated TM-like and TE-like modes can be spectrally close to each other and far from other modes. In this case, the system can be effectively described by a four-by-four non-Hermitian Hamiltonian (at the Γ point in the Brillouin zone):

$$H_{sym} = \begin{bmatrix} \omega^1_{TE} + i\gamma^1_{TE} & 0 & 0 & 0 \\ 0 & \omega^1_{TM} + i\gamma^1_{TM} & 0 & 0 \\ 0 & 0 & \omega^2_{TE} + i\gamma^2_{TE} & 0 \\ 0 & 0 & 0 & \omega^2_{TM} + i\gamma^2_{TM} \end{bmatrix} \quad (1)$$

with $\omega^i_{TE(TM)}$ and $\gamma^i_{TE(TM)}$ ($i=1, 2$) being the resonant frequency and the decay rate of the degenerated TE-like (TM-like) modes. Moreover, these parameters satisfy the relationship of $\omega^1_{TE(TM)} = \omega^2_{TE(TM)}$ and $\gamma^1_{TE(TM)} = \gamma^2_{TE(TM)}$.

By breaking the up-down symmetry ($n_b \neq n_s$), the orthogonal TE-like and TM-like modes can couple with each other. In this case, the effective-Hamiltonian of the system can be written as:

$$H_{asym} = \begin{bmatrix} \omega^1_{TE} + i\gamma^1_{TE} & \kappa & 0 & 0 \\ \kappa^* & \omega^1_{TM} + i\gamma^1_{TM} & 0 & 0 \\ 0 & 0 & \omega^2_{TE} + i\gamma^2_{TE} & \kappa \\ 0 & 0 & \kappa^* & \omega^2_{TM} + i\gamma^2_{TM} \end{bmatrix}, \quad (2)$$

where $\kappa$ quantifies the coupling strength which can be tuned by changing the background refractive index $n_b$ to control the breaking degree of the up-down symmetry. In this case, the interaction Hamiltonian [0 $\kappa$; $\kappa^*$, 0] is a Hermitian operator because the interaction between the TE-like and TM-like modes is caused mainly by evanescent field instead of far-field coupling via the radiation continuum. By turning the coupling strength between the orthogonal polarization modes and their radiation losses (through changing the geometric parameters of the system), the eigenvectors of $H_{asym}$ can coalesce, giving rise to one pair of degenerated exceptional points. Such EPs are very different from that existing in the symmetric photonic crystal slab described in Ref. [34], where only TE-like modes are coupled. Here, our proposed EP is formed by two orthogonal polarization modes in the asymmetric photonics crystal slab, it can be called as the radiation vector EP.

Fig. 1b and 1c show the real and imaginary parts of the band structure in the two-parameter space (each band is doubly degenerated) where the refractive index of the background and thickness of the photonic crystal slab are swept. The remained parameters are chosen to be $a=259$nm, $d=136$nm, $n_p=n(Si_3N_4)=2.02$ and $n_s=n(Si_2O_3)=1.47$, respectively. The eigenvalues of the system are calculated by eigenfrequency solver of COMSOL Multiphysics. It is clearly shown that both the real and imaginary parts of the eigenspectra possess the characteristics of a self-intersecting Riemann surface. The EP marks the branch point where the Riemann surface splits. In Fig. 1d, we further plot the real and imaginary parts of eigenvalues for the system as a function of the background refractive index when the thickness of the photonic crystal slab is fixed as $h$=154.2nm. It proves the existence of the EP when the background refractive index is set as $n_b$=1.3845 (both the real and imaginary parts of eigenvalue being identical). Except for the square-root dispersion around the EP, the other important feature of the EP is that the corresponding eigenstates should also be coalesced. In the S1 of the supporting information, we present the numerical results of eigenfield distribution with different background mediums. We found that the eigenfield distribution for the two hybridized eigenstates becomes more and more similar to each other with the system approaching the exceptional point, which demonstrates the coalesce of eigenvectors.

This vector EP possesses a potential application for the realization of strong and single-handed superchiral fields. Many previous works [20] have proved that the excitation of TE (or electric) and TM (or magnetic) modes with $\pi/2$ phase shift is necessary to achieve the spatially homogeneous superchiral fields (locally displaying greater optical chirality than CPLs). In this consideration, our proposed vector EP may become a suitable candidate for this purpose. This is due to the fact that the eigenmode of our designed vector EP [$\psi$(EP)] ($\psi$=[$E_x$ $E_y$ $E_z$ $H_x$ $H_y$ $H_z$]) can be expressed as $\psi$(EP)=$\psi$(TE)-$i\psi$(TM). Taking the $x$-polarized case for example, close to the central plane, the $\psi$(TE) and $\psi$(TM) take the forms of $\psi$(TE)=[$E_x^{(TE)}$ 0, 0, 0, 0, $H_z^{(TE)}$] and $\psi$(TM)=[0 0, $E_z^{(TM)}$, $H_x^{(TM)}$, 0, 0], respectively. Thus, at the EP, the modal electric fields possess a large component parallel to the magnetic fields. The blue and green arrows in Fig. 1(e) show the real part of electric fields and imaginary part of magnetic fields in a unit cell of the photonic crystal slab for the EP. At each location, the electric field has a large component parallel to the magnetic field. Such a phenomenon is absent if the mode is purely TE or TM polarized. From the definition C=-$\varepsilon_0\omega$Im($E^*$.B)/2, these parallel and 90-degree out-of-phased electric and magnetic fields lead to a uniform and single handed optical chirality field. Consequently, net superchiral fields may be produced by effectively exciting the photonic crystal slab at the vector EP. In this case, our designed vector EP can be used for ultra-sensitive chiral detection.

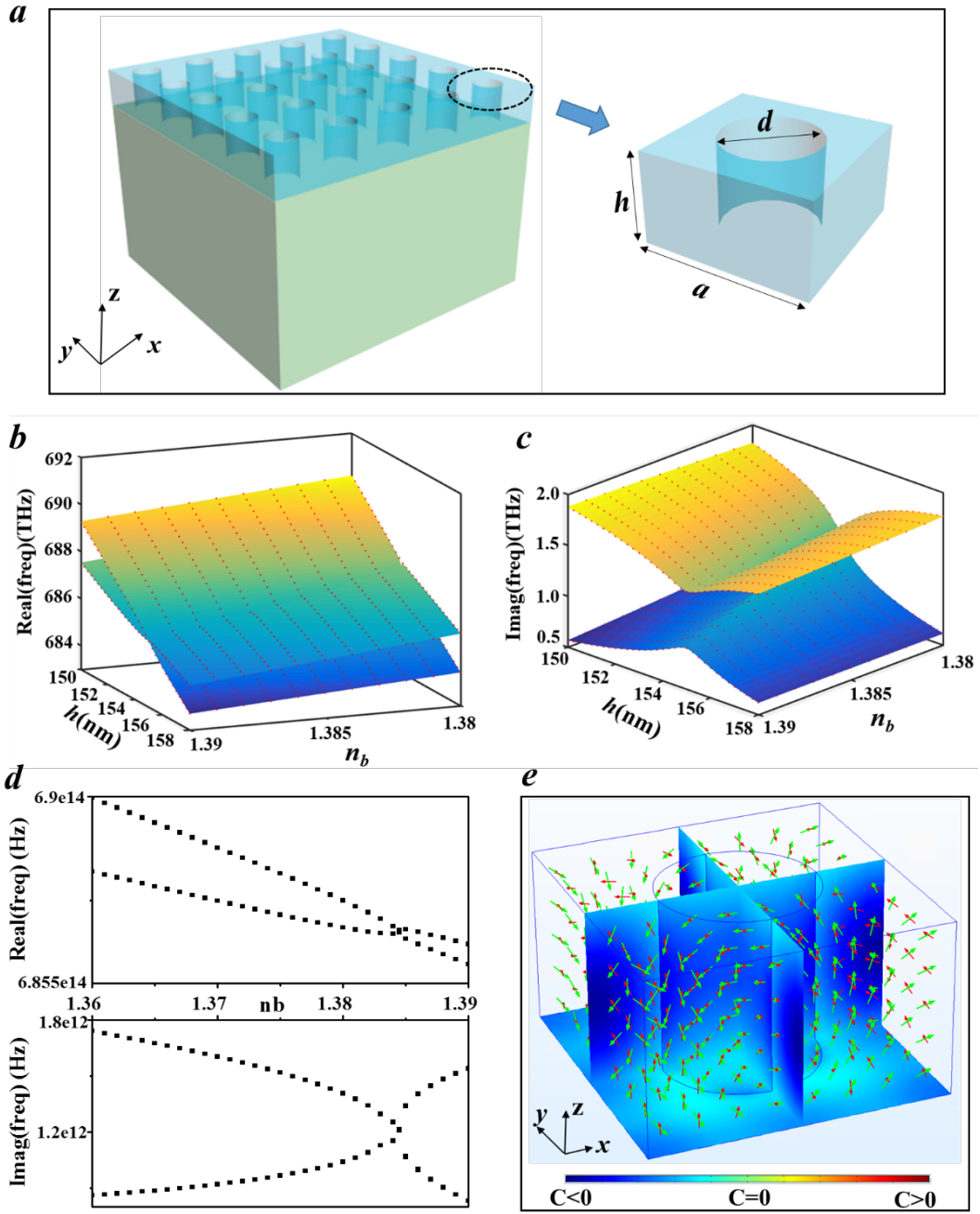

FIG. 1 (a) Diagram of the asymmetric photonics crystal slab. The real (b) and imaginary (c) parts of the eigenvalues for the system in the two parameters space, where the background refractive index and the thickness of the photonic crystal slab are swept. Other parameters are chosen as $a$=259nm, $d$=136nm, $n_p$=2.02 and $n_s$=1.47, respectively. (d) The real and imaginary parts of eigenvalues for the system as a function of the refractive index of the background medium when the thickness of the photonic crystal slab is fixed as $h$=154.2nm. (e) The optical chirality of the eigenmode at the vector EP. The red and green arrows correspond to the real and imaginary parts of the electric and magnetic fields.

To verify our theoretical design of the vector EP, we fabricate a $Si_3N_4$ photonic crystal slab on top of the silica substrate using electron beam lithography (see S2 in the supporting information for details). Scanning electron microscope images (both side and top views) of the sample are shown in Fig. 2a. The sample is about $50\times50um^2$ and the period of the square lattice is $a$=259nm. The thickness and diameter of cylindrical holes are $h$=154nm and $d$=136nm, respectively. These parameters are nearly identical with the structure sustaining the vector EP discussed above. The only difference is that the absorption loss of $Si_3N_4$ photonic crystal slab should be considered in the real experiments, where the refractive index of $Si_3N_4$ is chosen as $n(Si_3N_4)$=2.02+ $i$0.003. It is worthy to note that the vector EP still exists even the photonic crystal slab possesses small intrinsic losses (see Fig. S2a and S2b in the supporting information). Based on the eigenspectra calculation, we find that two modes can merge into one at the vector EP by increasing the background refractive index (shown in Fig. S2 and Fig. 1d). This special property can be illustrated through the calculated/measured transmission spectra.

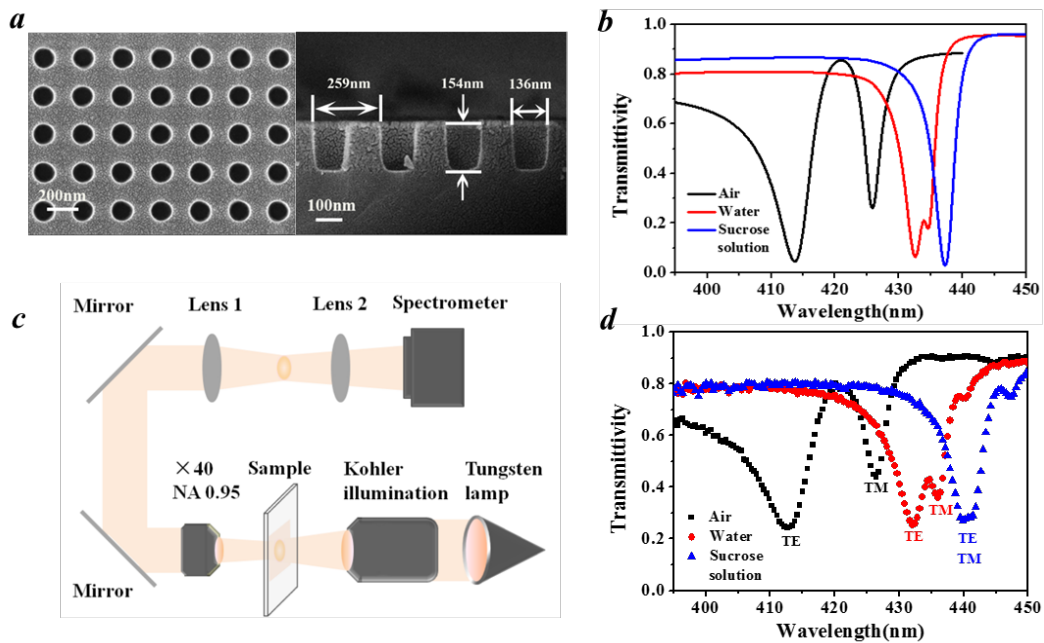

FIG. 2. (a) Side and top views of the scanning electron microscope images for the fabricated sample. The geometric parameters of the fabricated samples are $a$=259nm, $d$=136nm and $h$=154nm, respectively. (b) and (d) The calculated and measured transmission spectra of the structure possessing the vector EP. The black, red and blue lines/dots correspond to the conditions with the structure immersing into air, water and sucrose solution, respectively. (c) Experimental setups for the measurement of transmission spectra.

We firstly calculated the transmission spectra of the above structure, where the photonic crystal slab is immersed into various background media (air, water and sucrose solution). As shown in Fig. 2b, the black line marks the transmission spectrum with the background medium being air. Two separated valleys correspond to the excitation of TE-like and TM-like modes.

Due to the intrinsic loss of the $Si_3N_4$ photonic crystal slab, the transmittivities at these two valleys are not equal to zero. The red and blue lines correspond to cases where the background media are water and sucrose solutions, respectively. We found that the two separated valleys get much closer when the system is immersed into water ($n_b$=1.33). Finally, a single valley of the transmission spectrum appears when the refractive index of the background medium reaches to $n_b$=1.385. In this case, the vector EP is proved to be achieved.

To experimentally demonstrate the realization of the vector EP, the transmission spectra within different background media are measured by using the homemade experimental set-up [62], as shown in Fig. 2c. Details of measurement system are provided in the S2 part in the supporting information. The measured transmission spectra are plotted in Fig. 2d. We find that the two separated modes can coalesce into one when the concentration of sucrose solution reaches to 75% ($n_b$=1.385). The measured results with different concentration of sucrose solution are plotted in Fig. S3. This phenomenon possesses a good agreement with numerical results. The extra little dip (near 448nm in 75% sucrose solution), which should not exist ideally, may result from the defect mode caused by fabrication errors or non-vertical illumination of the system. The lower Q-factor (larger width of the transmission spectra) compared with the theoretical prediction may result from the finite size effect of samples (about $50\times50um^2$) and scattering losses caused by fabrication imperfections and disorders [63].

Although the eigenstate of our designed vector EP possesses single-handed optical chirality (shown in Fig. 1e), it is not a simple task to effectively excite it. Because the asymmetric structures always support different near-field distributions and electromagnetic responses when they are excited from different directions [64]. In this case, it is not sure how to excite the asymmetric photonic crystal slab to create the single-handed superchiral field at the vector EP.

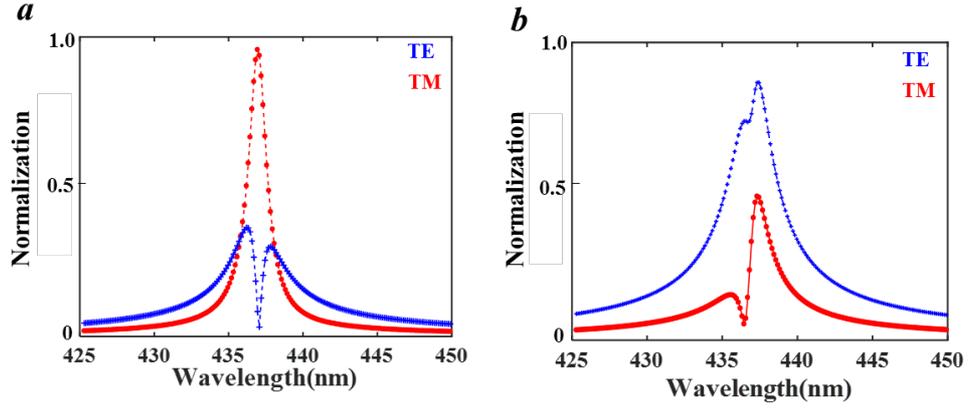

FIG. 3. The variation of absolute values for the normalized excitation strength of TE-like (blue dot) and TM-like (red dot) modes as functions of the wavelength for the incident light propagating along the +z (a) and –z (b) directions. The structural parameters are chosen as $a$=259nm, $d$=136nm, $n_p$= 2.02 and $n_s$=1.47, $n_b$=1.385 and $h$=154.2nm, where the vector EP exists in the system.

In order to explore the excitation property of TE-like and TM-like modes, we develop the temporal coupled-mode theory for the system with two resonance modes and a pair of radiation-channels (upward and downward ports). The analytical model is given in the section S4 of the supporting information. By using the proposed temporal coupled-mode theory, the modal excitation with different incident wavelengths and directions can be clearly analyzed. Here, the used parameters are identical with the structure (used in Fig. 1e) possessing the vector EP. As shown in Fig. 3a, we plot the variation of absolute values for the normalized excitation strength of TE-like (blue dot) and TM-like (red dot) modes as functions of the wavelength for the incident plane wave propagating along the +z-direction (incident from the substrate). It is clearly shown that the TM-like mode is mainly excited at the vector EP and the excitation strength for the TE-like mode is minimum in this case. In contrast, when the incident light is coming from -z-axis (incident from the background), the TE-like mode can be significantly excited and the TM-like mode is nearly not raised, as shown in Fig. 3b. Consequently, it is clearly shown that the TM-like (TE-like) mode is significantly coupled to the upward (downward) port at the vector EP. In this case, the homogeneous superchiral fields cannot be created nearby the photonic crystal slab excited by a single beam of plane wave coming from either upward port or downward port. Hence, a new excited method should be proposed to generate the single-handed superchiral field at the vector EP.

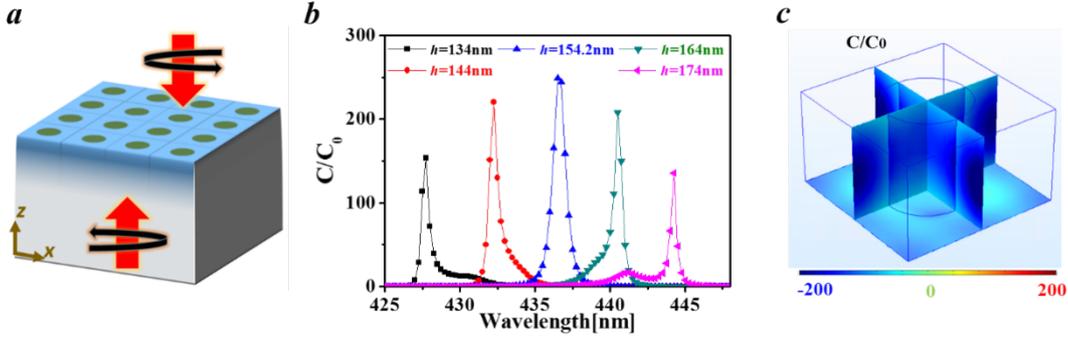

FIG. 4. (a) The scheme to generate the superchiral field at the vector EP by using two beams of CPL exciting the system from opposite directions. (b) Left chart: the averaged enhancement of $C/C_0$ in the cylindrical hole of the photonic crystal slab with different thickness. The blue line corresponds to the structure sustaining EP with $h$=154.2nm. The black, red, green, and pink lines correspond to cases deviating the vector EP. (c) Near-field distribution of optical chirality near the photonic crystal slab at the vector EP under two beams excitation from opposite directions.

Based on the modal-coupling property at the vector EP, we design a feasible scheme to generate the single-handed superchiral field at the vector EP by using two beams of CPLs to excite the system from opposite directions, as shown in Fig. 4a. Fig. 4b shows the averaged enhancement of $C/C_0$ in the cylindrical hole of the photonic crystal slab with different thicknesses. $C_0$ is the averaged optical chirality of the CPL. We find that the maximal enhancement of optical chirality appears in the system sustaining the vector EP (blue line, $h$=154.2nm). In this case, the corresponding distribution of enhancement factor for the optical chirality near the structure is plotted in Fig. 4c. It is clearly shown that single-handed superchiral field, which is consistent with the eigenstate at the vector EP (shown in Fig. 1e), is created. When the thickness of the photonic crystal slab deviated from that of structure possessing the vector EPs, the enhancement factor of $C/C_0$ is decreased, as shown in Fig. 4b (black, red, green, and pink lines). From the above results, we find that our designed photonic crystal slab sustaining the vector EP can generate the maximal near-field optical chirality under two beams excitation from opposite directions. Additionally, it is also important to note that the generated superchiral fields are also related to the initial relative phase of the two incident lights coming from opposite channels. This is owing to the fact that the relative phase between excited TE- and TM-like modes can also significantly influence the chirality of near-fields. Only $\pi/2$ phase shift between TE- and TM-like modes can induce spatially homogeneous superchiral fields (see the supporting information for details). With the utilization of such strong and single-handed superchiral fields, our designed photonic crystal slab can be used to perform surface-enhanced fluorescence CD and Raman optical activity of chiral molecules, which are proportional to the optical chirality at the molecular positions.

In conclusion, we propose theoretically and demonstrate experimentally a new type of EP, which is called radiation vector EP, to realize the strong superchiral fields for chiral molecular detection. By breaking the up-down symmetry, the TE-like and TM-like modes can couple with each other, and coalesce as the vector EP. With the help of temporal coupled-mode theory, we propose a scheme to generate single-handed superchiral field near the photonic crystal slab at the vector EP where two beams of lights are used to excite the system from opposite directions. By using generated strong and homogenous superchiral fields, our designed photonic crystal slab could be an ideal platform to perform surface-enhanced fluorescence CD and Raman optical activity of chiral molecules. Moreover, by using the concept proposed in our work and the higher-order EP, we could design nanodevices with much-higher sensitivity of chiral detection based on their resonance shifts. Some candidate systems include nanostructures with high-quality factors, such as the photonic crystal cavities, microspheres or microdisks. Our findings may provide a guideline for the design of novel chiral nanosensors for applications in the fields of biomedicine and pharmaceutics.

## ACKNOWLEDGEMENTS

This work was supported by the National key R & D Program of China under Grant No. 2017YFA0303800 and the National Natural Science Foundation of China (91850205 and 11574031).

[1] N. Berova, P. L. Polavarapu, K. Nakanishi, and W. Woody, *Comprehensive Chiroptical Spectroscopy* (Wiley, New York, 2012).
[2] Busch, K. W. and Busch, M. A, *Chiral Analysis* (Elsevier, Amsterdam, 2006).
[3] H. J. Rhee *et al.*, Nature **458**, 310 (2009).
[4] D. Patterson, M. Schnell, and J. M. Doyle, Nature **497**, 475 (2013).
[5] D. Sofikitis, L. Bougas, G. E. Katsoprinakis, A. K. Spiliotis, B. Loppinet, and T. P. Rakitxis, Nature **514**, 76 (2014).

[6] G. D. Fasman, *Circular dichroism and the conformational analysis of biomolecules* (Plenum, New York, 1996).

[7] G. H. Wagniere, *On chirality and the universal asymmetry: reflections on image and mirror image* (VHCA with Wiley-VCH, Zurich, 2007).

[8] B. M. Maoz,, Y. Chaikin, A. B. Tesler, O. B. Elli, Z. Fan, A. O. Govorov, and G. Markovich, Nano Lett. **13,** 1203 (2013).

[9] W. Ma, H. Kuang, L. Xu, L. Ding, C. Xu, L. Wang, N. A. Kotov, Nat. Commun. **4,** 2689 (2013).

[10] Y. Zhao, A. N. Askarpour, L. Sun, J. Shi, X. Li, and A. Alu, Nat. Commun., **8,** 14180 (2017).

[11] E. Mohammadi, K. L. Tsakmakidis, A. N. Askarpour, P. Dehkhoda, A. Tavakoli, H. Altug, ACS Photonics **5,** 2669 (2018).

[12] Y. Tang, A. E. Cohen, Phys. Rev. Lett. **104,** 163901 (2010).

[13] Y. Tang and A. E. Cohen, Science **332,** 333 (2011).

[14] C. He *et al.*, Nat. Commun. **10,** 2093 (2019).

[15] E. Hendry *et al.*, Nat. Nanotechnol. **5,** 783-717 (2010)

[16] M. Schaferling, D. Dregely, M. Hentschel, H. Giessen, Phys. Rev. X **2,** 031010 (2012).

[17] M. Schäferling, X. Yin, and H. Giessen, Opt. Express **20,** 26326 (2012)

[18] T. J. Davis and E. Hendry, *Phys. Rev. B* **87,** 085405 (2013).

[19] A. Vazquez-Guardado and D. Chanda, Phys. Rev. Lett. **120,** 137601 (2018).

[20] A. Garcia-Etxarri and J. A. Dionne, Phys. Rev. B, **87,** 235409 (2013).

[21] R. Y. Wang *et al.*, J. Phys. Chem. C **118,** 9690 (2014).

[22] R. Tullius *et al.*, J. Am. Chem. Soc. **137,** 8380 (2015).

[23] F. Graf, J. Feis, X. Garcia-Santiago, M. Wegener, C. Rockstuhl, and I. Fernandez-Corbaton, ACS. Photonics **6**, 482 (2019).

[24] T. Wu, J. Ren, R. Wang, X. Zhang, J. Phys. Chem. C **118,** 20529 (2014).

[25] W. Zhang, T. Wu, R. Wang, X. Zhang, J. Phys. Chem. C **121,** 666 (2017).

[26] C. M. Bender and S. Böttcher, Phys. Rev. Lett. **80,** 5243 (1998).

[27] N. Moiseyev, Cambridge Univ. Press, (2011).

[28] M. A. Miri and A. Alù, Science **363,** eaar7709 (2019).

[29] H. Cao and J. Wiersig, Rev. Mod. Phys. **87,** 61 (2015)

[30] B. Peng *et al.*, Nature Phys. **10,** 394 (2014).

[31] S. Klaiman, U. Günther, and N. Moiseyev, Phys. Rev. Lett. **101,** 080402 (2008).

[32] C. E. Ruter *et al.*, Nature Phys. **6,** 192 (2010).

[33] C. Hahn *et al.*, Nat. Commun. **7,** 12201 (2016).

[34] B. Zhen e*t al.*, Nature **525**, 354 (2015).

[35] Z. Lin, A. Pick, M. Lončˇar, and A. W. Rodriguez, Phys. Rev. Lett. **117,** 107402 (2016).

[36] K.-H. Kim *et al.*, Nat. Commun. **7, 1**3893 (2016).

[37] H. Zhou *et al.*, Science **359,** 1009 (2018).

[38 L. Chang *et al.*, Nat. Photon. **8,** 524 (2014).

[39] N. Caselli *et al.*, Nat. Commun. **9,** 396 (2018).

[40] Z. J. Wong *et al.*, Nat. Photon. **10,** 796 (2016).

[41] B. Peng *et al.*, Science **346,** 328–332 (2014).

[42] M. Liertzer *et al.*, Phys. Rev. Lett. **108,** 173901 (2012).

[43 K. G. Makris *et al.*, Phys. Rev. Lett. **100,** 103904 (2008).

[44] A. Guo *et al.*, Phys. Rev. Lett. 103, 093902 (2009).

[45] Y. Chong, L. Ge, and A. D. Stone, Phys. Rev. Lett. **106,** 093902 (2011).

[46] Z. Lin *et al.*, Phys. Rev. Lett. **106,** 213901 (2011).

[47] A. Regensburger *et al.*, Nature **488,** 167 (2012).

[48] J. W. Yoon *et al.*, Nature **562**, 86 (2018).

[49] J. Doppler *et al*., Nature **537,** 76 (2016).

[50] H. Xu, D. Mason, L. Jiang, and J. G. E. Harris, Nature **537,** 80 (2016).

[51] C. Dembowski *et al.*, Phys. Rev. Lett. **86,** 787 (2001).

[52] H. Hodaei, M.-A. Miri, M. Heinrich, D. N. Christodoulides, and M. Khajavikhan, Science **346,** 975 (2014).

[53] L. Feng, Z. J. Wong, R.-M. Ma, Y. Wang, and X. Zhang, Science **346**, 972 (2014)

[54] E. Lafalce *et al.*, Nat. Commun. **10,** 561 (2019).

[55] J. Wiersig, Phys. Rev. Lett. **112,** 203901 (2014).

[56] G. Zhang, Y. Wang and J. You, Phys. Rev. A **99**, 052341 (2019).

[57] Z.-P. Liu *et al*., Phys. Rev. Lett. **117,** 110802 (2016).

[58] W. Chen, Özdemir, S. K., Zhao, G., Wiersig, J. & Yang, L. Nature **548,** 192 (2017).

[59] H. Hodaei *et al.*, Nature **548,** 187–191 (2017).

[60] H. K. Lau and A. A. Clerk, Nat. Commun. **9,** 4320 (2018).

[61] S. Wang *et al.*, Nat. Commun. **10**, 832 (2019).

[62] J. C. Jin *et al.*, Nature **574**, 501 (2019).

[63] D. Sounas and A. Alù, Phys. Rev. Lett. **118,** 154302 (2017).

[64] Y. W. Zhang, A. Chen, W. Z. Liu, *et al.*, Phys. Rev. Lett. **120,** 186103 (2018).

# Supporting Materials

# Vector exceptional points with strong superchiral fields

Tong Wu[1*], Weixuan Zhang[1*], Huizhen Zhang[1*], Saisai Hou[1], Guangyuan Chen[1], Ruibin Liu[1], Cuicui Lu[1], Jiafang Li[1], Rongyao Wang[1], Pengfei Duan[2], Junjie Li[3], Lei Shi[4], Jian Zi[4] and Xiangdong Zhang[1$]

[1]Key Laboratory of advanced optoelectronic quantum architecture and measurements of Ministry of Education, Beijing Key Laboratory of Nanophotonics & Ultrafine Optoelectronic Systems, School of Physics, Beijing Institute of Technology, 100081, Beijing, China;

[2]CAS Center for Excellence in Nanoscience, CAS Key Laboratory of Nanosystem and Hierarchical Fabrication, National Center for Nanoscience and Technology (NCNST), Beijing 100190, China.

[3]Beijing National Laboratory for Condensed Matter Physics, Institute of Physics, Chinese Academy of Sciences, Beijing 100190, China

[4]State Key Laboratory of Surface Physics, Key Laboratory of Micro- and Nano-Photonic Structures (Ministry of Education) and Department of Physics, Fudan University, Shanghai 200433, China

*These authors contributed equally to this work.

[$]Author to whom any correspondence should be addressed: zhangxd@bit.edu.cn

## S1. The evolution of eigenfields with system approaching the EP.

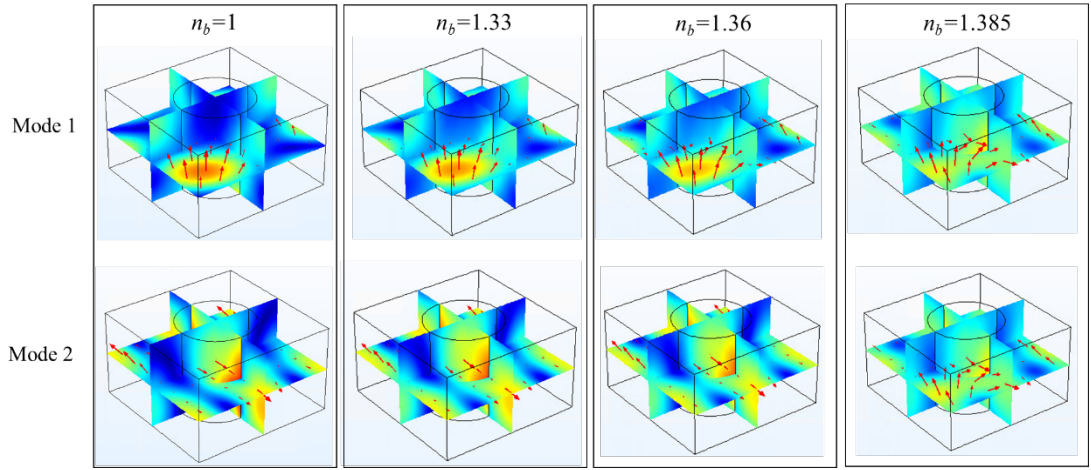

Figure S1. The distribution ($|E|^2$) and polarization (red arrows) of the eigenfields with different values of the background refractive index.

In this part, we present the evolution of eigenstates of our proposed PhC slab to further identify the occurrence of the EP induced by increasing the refractive index of the background medium. Except for the square-root dispersion around the EP, the other important feature of the

EP is that the corresponding eigenstates should also be coalesced. As shown in Fig. S1, we numerically calculated the distribution ($|E|^2$) and polarization (red arrows) of the eigenfield with different values of the background refractive index ($n_b$=1, 1.33, 1.36 1,385). It is noted that with the system approaching the exceptional point, the eigenfield distribution for the two hybridized eigenstates becomes more and more similar to each other. In addition, near the EP, the electric fields at the central-plane of the PhC slab possess both in-plane and out-of-plane components. These results clearly validate the interaction and hybridization between the two modes and demonstrate the coalesce of eigenvectors.

**S2. Sample fabrication and transmission spectrum measurement.**

The $Si_3N_4$ layer was grown with the plasma enhanced chemical vapor deposition method on a 500um-thick silica substrate, and subsequently coated with a layer of 300-nm polymethyl methacrylate (PMMA) as e-beam photoresist, a thin layer of 60-nm conductive coating to increase conductivity of the sample. The periodic square lattice pattern was created with electron-beam lithography (Zeiss SUPRATM 55 scanning electron microscopy plus Raith Elphy Quantum). After exposures, the sample was developed and the pattern was transferred from PMMA to $Si_3N_4$ by reactive-ion etching (Oxford NGP80); CF4/CHF3/O2 gas was used to etch $Si_3N_4$ with PMMA be used as etching mask.

The transmission spectrum was measured using a home-made imaging spectroscopy built based on an Olympus microscope (IX73). The source is a tungsten lamp and the incident light is focused on to sample by an objective (40_ magnitude, NA 0.95). A schematic view of the setup is shown in Fig. 2b. The transmitted light passed through sample was imaged onto the entrance slit of imaging spectrometer (Princeton Instruments IsoPlane-320) using a series of convex lens. Thus the transmission light comes from different angles have corresponding position on the entrance slit. In our experiment, the transmission light perpendicular to sample was selected, in other words, transmission spectrum of incident light perpendicular to sample was selected.

**S3. The influence of intrinsic absorption of $Si_3N_4$ on the vector EP.**

Here, we prove that the small value of the intrinsic absorption of $Si_3N_4$ nearly has negligible influence on the formation of the vector EP. In Fig. S2a and S2b, we plot the real and imaginary parts of eigenvalues of the asymmetric photonic crystal slab as functions of the refractive index of the background medium. The system parameters are identical with the structure used in Fig. 2d except for the refractive index of $Si_3N_4$ being $n(Si_3N_4)=2.02+0.003i$. We note that our designed system still possesses the vector EP even intrinsic losses exist. The only difference is that the imaginary part of eigenvalues gets larger. We also calculated the eigenspectra of the system with different lossy values for the $Si_3N_4$ medium. Figs. S2c and S2d (Figs. S2e and S2f) present the numerical results of the real and imaginary parts of the eigenspectra when the refractive index of $Si_3N_4$ is selected as $n(Si_3N_4)=2.02+0.004i$ [$n(Si_3N_4)=2.02+0.005i$], respectively. We find that the system deviates a little from the ideal vector EP with larger lossy values.

From the about result, we note that the vector EP can still exists with small intrinsic absorptions. With the absorption being increased, the system may deviate from the ideal vector EP.

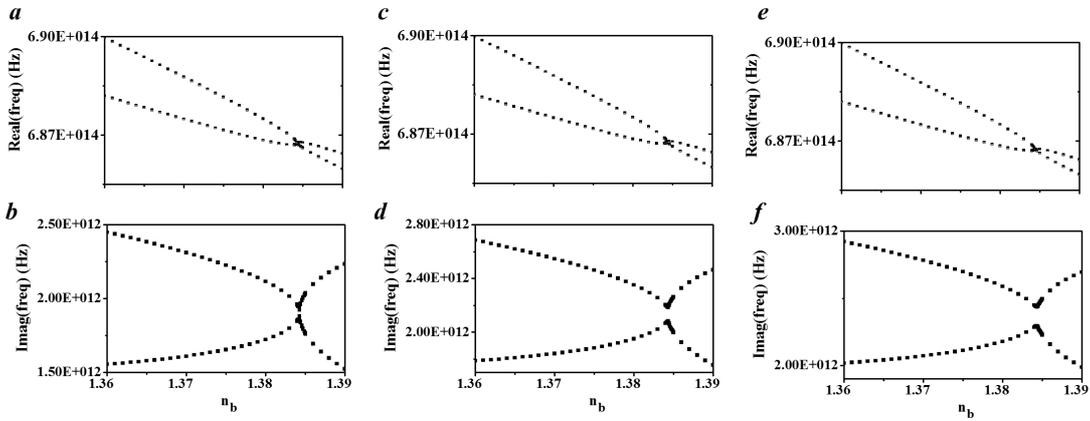

Figure S2. The real and imaginary parts of eigenvalues for the system with intrinsic losses. The refractive index of $Si_3N_4$ is chosen as $n=2.02+0.003i$ for (a) and (b), $n=2.02+0.004i$ for (c) and (d) and $n=2.02+0.005i$ for (e) and (f).

**S4. The measured transmission spectra of the structure possessing vector EP.**

In this part, we add some measured results with different concentrations of sucrose (25%, 50% and 75%), as plotted in Fig. S3. It is clearly shown that the two separated valleys gradually get closer to each other by increasing the refractive index of the background media. And, these two modes finally merge into one when the concentrations of sucrose reaches to 75%, indicating the formation of the vector EP.

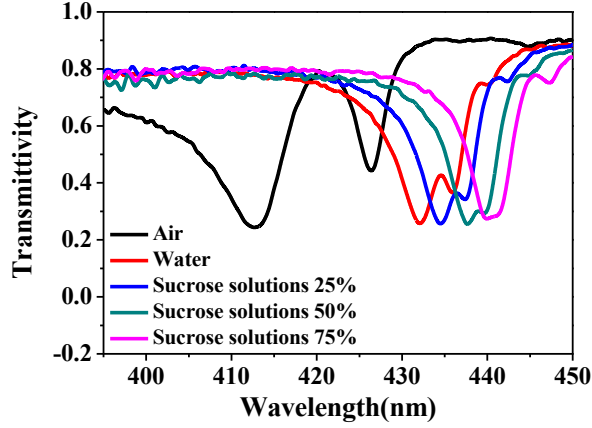

**Figure S3** Measured transmission spectra of the structure.

## S5. Temporal coupled-mode theory of the asymmetric photonic crystal slab.

In order to explore the excitation property of TE-like and TM-like modes, we develop the temporal coupled-mode theory for the asymmetric PCS with two resonance modes (TE-like and TM-like modes) and two radiation-channels (upward port and downward port). Due to the fact that the doubly degenerated TE-like (or TM-like) modes are decoupled from each other (protected by the $C_{4v}$ symmetry), in this case, only one pair of TE- and TM-liked modes are able to describe the spectral response of the system.

Based on time reversal symmetry, the dynamics of the amplitudes of TE-like and TM-like modes can be described by:

$$\frac{d}{dt}\begin{pmatrix} A_{TE} \\ A_{TM} \end{pmatrix} = i\begin{bmatrix} \omega_{TE} & \kappa \\ \kappa & \omega_{TM} \end{bmatrix}\begin{pmatrix} A_{TE} \\ A_{TM} \end{pmatrix} - \begin{bmatrix} \gamma_{TE} & 0 \\ 0 & \gamma_{TM} \end{bmatrix}\begin{pmatrix} A_{TE} \\ A_{TM} \end{pmatrix} + \begin{bmatrix} d_{TE} & d_{TE} \\ -d_{TM} & d_{TM} \end{bmatrix}\begin{pmatrix} s_{1+} \\ s_{2+} \end{pmatrix} \quad (S1)$$

$$\begin{pmatrix} s_{1-} \\ s_{2-} \end{pmatrix} = \exp(i\phi)\begin{bmatrix} R & iT \\ iT & R \end{bmatrix}\begin{pmatrix} s_{1+} \\ s_{2+} \end{pmatrix} + \begin{bmatrix} d_{TE} & -d_{TM} \\ d_{TE} & d_{TM} \end{bmatrix}\begin{pmatrix} A_{TE} \\ A_{TM} \end{pmatrix} \quad (S2)$$

where $A_{TE}$ and $A_{TM}$ corresponds the amplitude of TE-like and TM-like modes. $\omega_{TE}$ and $\omega_{TM}$ are the center frequencies of the TE-like and TM-like resonances. $\gamma_{TE}$ and $\gamma_{TM}$ are the decay rates, which result from the radiation loss, of TE-like and TM-like modes. $S_{1+}$ and $S_{2+}$ are the amplitudes of the incoming wave from two radiation channels. $S_{1-}$ and $S_{2-}$ are the amplitudes of the outgoing waves toward two radiation channels. $\kappa$ is the evanescent coupling coefficient between the TE-like and TM-like modes induced by the substrate. $d_{TE}$ ($-d_{TM}$) is the coupling coefficient between TE-like (TM-like) mode with output port $\sim S_{2+}$. It is worthy to note that TE-like (TM-like) mode is the symmetric (asymmetric) mode about the middle-tangent plane. Hence, the coupling coefficients between the TE-like (TM-like) mode and two radiation channels have the same (opposed) sign. $R$ and $T$ are the real parts of the direct reflection and transmission coefficients. $\Phi$ is the corresponding phase factor depending on the positions of the reference plane.

According to the reciprocity and time reversal symmetry of the system, $d_{TE}$ and $d_{TM}$ can be expressed as:

$$\begin{aligned} d_{TE} &= \sqrt{\gamma_{TE}}\sqrt{-\exp(i\phi)[R+iT]} \\ d_{TM} &= \sqrt{\gamma_{TM}}\sqrt{-\exp(i\phi)[R-iT]} \end{aligned} \tag{S3}$$

When the angular frequency of the incident light is $\omega$, the Eqs. S1 and S2 can be reduced as:

$$\begin{pmatrix} A_{TE} \\ A_{TM} \end{pmatrix} = V \begin{pmatrix} s_{1+} \\ s_{2+} \end{pmatrix} \tag{S4}$$

$$\begin{pmatrix} s_{1-} \\ s_{2-} \end{pmatrix} = G \begin{pmatrix} s_{1+} \\ s_{2+} \end{pmatrix} \tag{S5}$$

with the matrix $V$ and $G$ being expressed as:

$$V = -i \begin{bmatrix} \omega - \omega_{TE} - i\gamma_{TE} & -\kappa \\ -\kappa & \omega - \omega_{TM} - i\gamma_{TM} \end{bmatrix}^{-1} \begin{bmatrix} d_{TE} & -d_{TM} \\ d_{TE} & d_{TM} \end{bmatrix}^{T} \tag{S6}$$

$$G = \exp(i\phi)\begin{bmatrix} R & iT \\ iT & R \end{bmatrix} - i\begin{bmatrix} d_{TE} & -d_{TM} \\ d_{TE} & d_{TM} \end{bmatrix}\begin{bmatrix} \omega - \omega_{TE} - i\gamma_{TE} & -\kappa \\ -\kappa & \omega - \omega_{TM} - i\gamma_{TM} \end{bmatrix}^{-1}\begin{bmatrix} d_{TE} & -d_{TM} \\ d_{TE} & d_{TM} \end{bmatrix}^{T} \tag{S7}$$

Firstly, we focus on the case with the incident light coming from the downward port (along

+z-axis, $S_{1+}\neq 0$ and $S_{2+}=0$). Based on the Eq. S5, the reflection coefficient should be expressed as: $S_{1-}/S_{1+}=G_{11}$. By fitting the numerically calculated reflection spectrum (using COMSOL) with this formula, the value of the remained parameters $\omega_{TE}$, $\omega_{TM}$, $\kappa$, $\gamma_{TE}$, $\gamma_{TM}$ can be obtained.

On the other hand, based on the Eq. S4, the amplitude of TE-like and TM-like modes can be described as: $A_{TE}=V_{11}S_{1+}$ and $A_{TM}=V_{21}S_{1+}$. In this case, we can numerically calculate the modal excitation strength with the fitting parameters $\omega_{TE}$, $\omega_{TM}$, $\kappa$, $\gamma_{TE}$, $\gamma_{TM}$, as shown in Fig. 3c.

Similarly, when the incident light comes from the upward radiation-channel (along -z-axis, $S_{1+}=0$ and $S_{2+}\neq 0$), the reflection coefficient should be expressed as: $S_{2-}/S_{2+}=V_{22}$. And, the amplitude of TE-like and TM-like modes are $A_{TE}=G_{12}S_{2+}$ and $A_{TM}=G_{22}S_{2+}$, respectively. By using the same method, the modal excitation strength in this case can be obtained, as plotted in Fig. 3d.

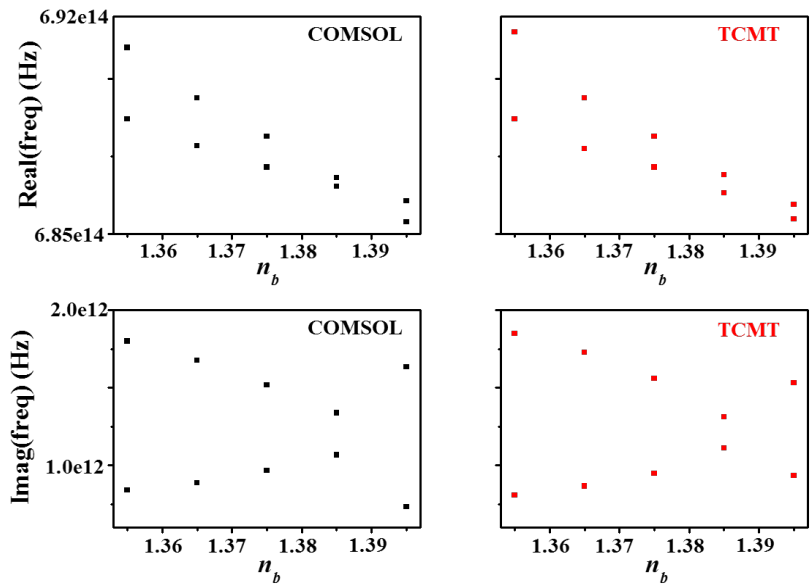

**Figure S4.** The comparison of eigenvalues fitted by temporal coupled-mode theory and calculated by COMSOL Multiphysics.

To prove the validity of the temporal coupled mode theory, in the following, we fit the numerically calculated transmission spectrum of the PhC slab with the formula $S_{1-}/S_{1+}=G_{11}$. In this case, the eigenvalues of TE-like ($\omega_{TE}$) and TM-like ($\omega_{TM}$) modes can be obtained. Here, we chose $a$=259nm, $d$=136nm, $h$=154.2nm, $n_p$=2.02 and $n_s$=1.47, respectively. The background refractive index varies from 1.36 to 1.39. As shown in Fig. S4, both real and imaginary parts of eigenvalues calculated by COMSOL and temporal coupled-mode theory are nearly consistent with each other. This indicates the correctness of the proposed temporal coupled-mode theory.

## S6. The influence of the initial relative phase of the incident light coming from two channels on the excitation of the vector EP.

In this part, we demonstrate that the enhancement of $C/C_0$ is related to the initial relative phase (ϕ) of the incident light coming from two channels. As shown in Fig. S5a-5f, we plot the averaged enhancement of the local optical chirality ($C/C_0$) within the cylindrical hole with different relative phase between two incident waves coming from top and bottom ports. Here, ϕ is ranging from 0deg to 315deg with the equal interval. It is clearly shown that different enhancements of the optical chirality appears with different values of the initial relative phase. Only with appropriate relative phase, the vector EP can be effectively excited where the large local optical chirality ($C/C_0$) can be created. This is due to the fact that the relative phase between TE-like and TM-like modes at the vector EP should be $0.5\pi$ to induce the spatially homogeneous superchiral fields. In this case, we can suitably tune the relative phase between two incident waves from top and bottom ports to effectively excite the vector EP.

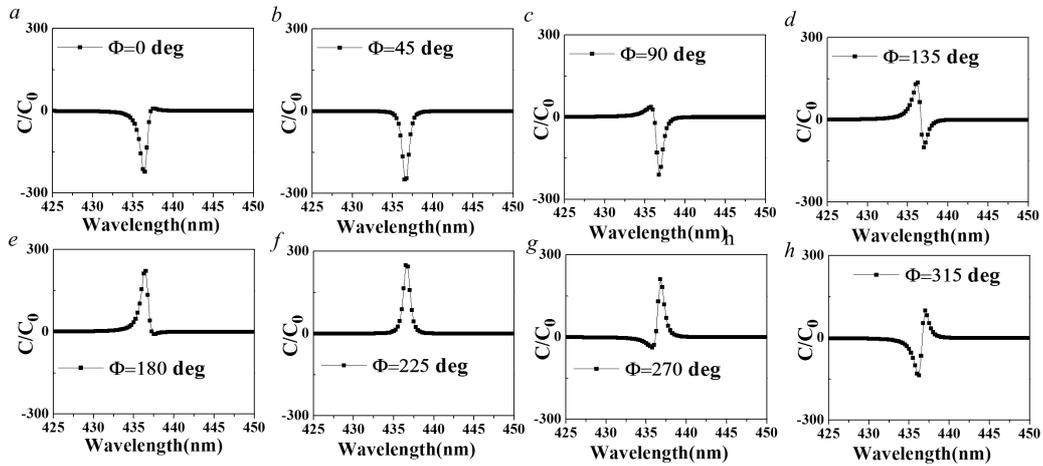

**Figure S5.** The averaged enhancement of the local optical chirality ($C/C_0$) within the cylindrical hole with different relative phase between two incident waves coming from top and bottom ports. The initial relative ϕ is equal to 0deg for (a), 45deg for (b), 90deg for (c), 135deg for (d), 180deg for (e), 225deg for (f), 270deg for (g) and 315deg for (h).